# A Tutorial on Modeling and Control of Slippage in Wheeled Mobile Robots

Khuram Naveed
Electrical Engineering Department
COMSATS Institute of Information Technology (CIIT)
Islamabad, Pakistan
Khurram.naveed@comsats.edu.pk

*Abstract*— **This tutorial intends to present a deep insight of the slip and skid problem in WMRs. Specifically, we present a brief introduction to slippage and the limitations it imposes on WMR's applications. In addition, we shed some light on state of the art review on modeling methodologies and controlling mechanisms for slip and skid in WMRs. This paper also discusses the future trends in the use of WMRs and possibility of new domains in their use in modern world.**

## I. INTRODUCTION

As the role of technology in our lives is increasing, robotic applications facilitating our daily lives are also increasing. Most of these applications demand a robot to be mobile. Wheels are one of the most efficient strategies for the motion of mobile robots because they need minimal control effort due to their inherent stability. In addition, they are simple in structure, consume less energy and faster than any other locomotion strategy. The applications of Wheeled Mobile Robots (WMRs) are increasingly present in industrial, commercial and defense activities. WMRs are potentially used for home assistance, transportation, mining, entertainment, surveillance and security today. Tomorrow WMRs will be extensively used for assistance in war zones and planetary explorations. Such vast scale of applications has made mobile robots very important control problem today and tomorrow. However the limitation of WMRs is their tendency to slip or skid on slippery, declining and uneven surfaces. These limitations have restricted the applications of WMRs on rough terrains and off the road explorations. The dream of watching WMRs move on oily surface, wet course, sand paths and icy terrain can only be fulfilled by minimizing slippage (slip and skid) in mobile robots.

From the last decade or so the control of WMRs has become an important problem mainly due to scope of their use in future. Earlier the nonholonomic constraints were considered for the control of mobile robots. Nonholonomic constraints dictate that wheel undergoes no slip and skid. This consideration however is far from reality as the performance of such control mechanisms is very poor in practical scenarios due to the presence of slip and skid. Different control strategies considering nonholonomic constraint are presented in [1-5].

In order to tackle this problem researchers included slip and skid dynamics in the robot model and then suitable control mechanisms for these models were developed [6], [7]. However different tasks require controlling and reducing slippage in WMRs i.e. motion control, stabilization control, trajectory tracking control, formation control etc. For all of these tasks different techniques are used for derivation of the model and designing control. In this paper we will present a brief overview of slip & skid and their modeling and control techniques.

This paper is divided in to five sections. Section I gives the preface of WMRs and their slippage problem while Section II provides the introduction to slip and skid. In Section III the detailed discussion on a few mathematical models of WMRs involving slippage has been presented. Section IV presents the state of the art of slip and skid control and Section V presents the concluding remarks on the slip control problem of WMRs.

## II. SLIPPAGE

Wheeled Mobile Robots are considered to be nonholonomic with their wheels experiencing no slippage under pure rolling. The controllers which are designed under the assumptions of no slip give poor tracking performance due to path deviation. This signifies the role of slip in WMRs. Thus for precise control it is important to model wheel slippage as a part of the model of a WMR. There are two types of slippage i.e. slip in the direction of motion of wheels and slip perpendicular to the direction of motion of wheel, the earlier is termed as longitudinal slip or simply 'slip' and latter is called lateral slip or 'skid'.

### A. Slip

This type of slip happens when the vehicle velocity does not remain equal to wheel's linear velocity [8]. An example of such a scenario is when brakes are applied to a high speed car where the speed of rotation of wheel is reduced suddenly decreasing its linear velocity. Whereas the body of the car wants to move at the same speed due to its momentum, thus creating a difference in the linear velocities of wheel and the car itself to generate slip. Another reason of slippage is tire deformation because it decreases angular velocity of the tire and hence decreasing its linear velocity compare to vehicle velocity. Slip may also be caused by slippery or oily surfaces. Thus following condition has to be satisfied to avoid slip.

$$v_v = v_w$$

Where, $v_v$ is the vehicle velocity and $v_w$ is the linear velocity of wheel. If $\dot{\varphi}$ is the angular velocity of the wheel and $r$ is the radius of the wheel then

$$v_w = r\dot{\varphi}$$

Thus for no slippage

$$v_v = r\dot{\varphi} \quad (1)$$

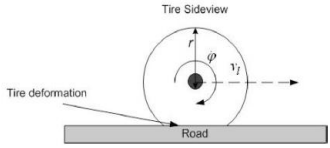

Fig. 1: Slipping due to tire deformation while its motion [9]

Thus the velocity responsible for the slip or the slip velocity $d$ is the difference between vehicle velocity and wheel velocity, mathematically,

$$d = v_v - r\dot{\varphi} \quad (2)$$

Slip ratio $\lambda$ can be defined as

$$\lambda = \frac{v_v - r\dot{\varphi}}{\max(v_v, r\dot{\varphi})} \quad (3)$$

This model for slip ratio implies $\lambda \in (0,1)$.

### B. Skid

While turning on the road a wheel experiences a lateral force at its contact patch with road forcing it away from its plan. This movement of the wheel away from its plan is called skid or lateral slip and the angle it moved away from its plan is called skidding angle $\delta$ [9]. The force responsible for skidding phenomena can be termed as centripetal force. Thus the value of the force responsible for skidding can be determined as $F = \cos\delta$.

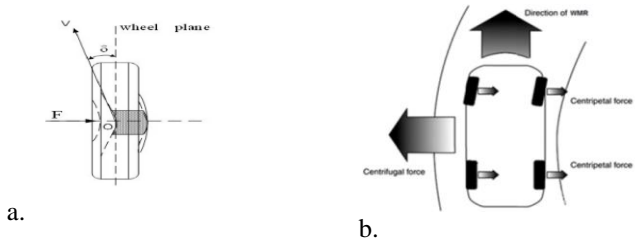

a.  
b.  
Fig. 2: Skidding phenomenon due to cornering effect of WMR [9].

## III. SLIP AND SKID MODELING IN WMRs

It has been already discussed that the control of WMRs under the nonholonomic constraints (i.e. no slip or skid) has not given the desired results due to the presence of slippage in real world scenarios. Thus to facilitate this problem slip models for WMRs were developed. In broader sense there are two approaches used in literature for modeling slip i.e. kinematic modeling involving slippage and road-tire interaction model to estimate friction. The earlier method exploits the kinematic constraints of a wheeled robot while the later studies the effect of forces on wheel while interacting with ground. Here we will discuss both of these approaches in detail.

### A. Modeling of slip through kinematics constraints

During the motion of a conventional wheel in a vertical plane, the wheel rotates along horizontal axle with changeable orientation. The contact between wheel and ground is assumed to be a single point. Kinematic constraints of a WMR dictate that the velocity of the material point of the wheel in contact with the ground is equal to zero [8].

Wheel slip is modeled using slip velocity $d = v_v - r\dot{\varphi}$, while slip angle $\delta$ is used to describe skidding phenomena in the kinematic model. Here we discuss the kinematic model of type (2,0) WMR presented in [10]. As shown in Fig. 3(b) a WMR is placed in frame $(X_b, Y_b)$ in global coordinates $(X, Y)$. The location of the center point of the robot P is described as $\xi = (x, y, \theta)$ where x and y are the co-ordinates of robots in XY-plane and $\theta$ gives the orientation of the robot w.r.t $(X_b, Y_b)$ in XY-plan. Velocity of the point P on robot is given by $v$ while its component along $X_b$ and $Y_b$ are $v_{xb}$ and $v_{yb}$ respectively. The relation between the slip angle $\delta$ and the horizontal and vertical components of robot velocity can be obtained from triangle in Fig. 3(a) below.

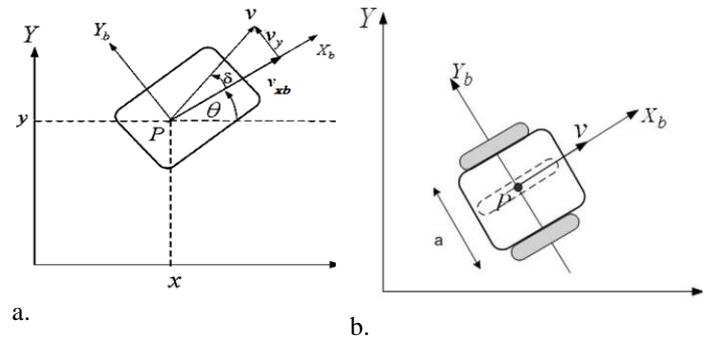

a.  
b.  
Fig. 3: Type (2,0) WMR posture and kinematics [10]

$$\tan\delta = \frac{v_{yb}}{v_{xb}} \quad (4)$$

The rotation matrix to express robot location in global co-ordinates from its local co-ordinates is given by

$$R(\theta) = \begin{bmatrix} \cos\theta & \sin\theta \\ -\sin\theta & \cos\theta \end{bmatrix} \quad (5)$$

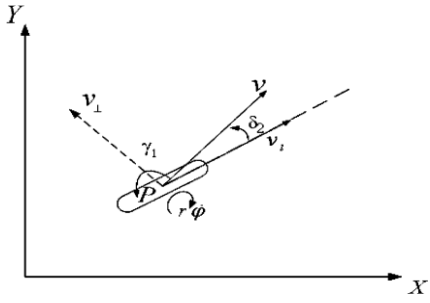

Fig. 4: Type (2,0) WMR kinematics with slippage [10].

The type (2,0) WMR shown in the Fig 3(b) has two wheels connected to the same axle, these two wheels may be represented by one fictitious wheel at the center of both the wheels. The Fig. 4 shows the motion of type (2,0) robot in the presence of slip and skid. When there is no slipping the velocity has a component along the direction of motion of the wheel, with vertical component equals to zero. But in the case of slippage the velocity deviates with an angle $\delta$ having a nonzero lateral component along $v_\perp$ and longitudinal component too along the direction of motion of wheel originally.

Thus the kinematic constraints involving slip and skid may be driven from the ideal case of pure rolling and no slipping in [11]. In case the condition of no slip is violated wheel velocity deviates an angle $\delta$, so the kinematic constraints now become

$$[-\sin\delta \quad \cos\delta]R(\theta)\dot{\xi} = [0] \quad (5)$$

The solution to this equation may be the longitudinal and lateral components of velocity in global frame.

$$\dot{\xi} = R(\theta)\begin{bmatrix} v\cos\delta \\ v\sin\delta \end{bmatrix}$$

$$\dot{\xi} = R(\theta)\begin{bmatrix} v_{xb} \\ v_{yb} \end{bmatrix}$$

$$\dot{\xi} = \begin{bmatrix} v_{xb}\cos\theta - v_{yb}\sin\theta \\ v_{xb}\sin\theta + v_{yb}\cos\theta \end{bmatrix} \quad (6)$$

Here $v_{yb} = v\sin\delta$ is the component of the velocity with which the wheel moves away from its plane representing the skidding phenomenon in the above kinematic model. While $v_{xb}$ denote the slipping in the model as when this velocity is less the $r\dot{\varphi}$ then wheel will slip. The yaw rate of the WMR can be interpreted as the sum of the input yaw rate $\gamma_1$ and the yaw rate of the WMR $\delta_1$ due to the wheel deviation because of slippage.

$$\dot{\theta} = \gamma_1 + \delta_1 \quad (7)$$

For no slippage

$$\dot{\theta} = \gamma_1$$

Thus the finally, the kinematic model involving slipping and skidding shown in eq. (8)

$$\dot{\xi} = \begin{bmatrix} v_{xb}\cos\theta + v_{yb}\sin\theta \\ v_{xb}\sin\theta + v_{yb}\cos\theta \\ \gamma_1 + \delta_1 \end{bmatrix} \quad (8)$$

Table below shows the kinematic models for slippage for few other configurations of WMRs. Fig. 5 shows the configurations all the robot models discussed the table below

KINEMATIC MODELS FOR SLIP AND SKID FOR DIFFERENT TYPE OF WMRs [10]

| Kinematic model with Slippage |
|---|
| Type(1,1) |
| $\dot{\xi} = \begin{bmatrix} v_{xb}\cos\theta + v_{yb}\sin\theta \\ v_{xb}\sin\theta + v_{yb}\cos\theta \\ \dfrac{v_{xb}}{C}\tan(\gamma_1 + \delta_1) - \dfrac{v_{yb}}{a} \end{bmatrix}$ |
| Type(2,1) |
| $\dot{\xi} = \begin{bmatrix} v\cos(\theta + \gamma_2 + \delta) \\ v\sin(\theta + \gamma_2 + \delta) \\ \dfrac{v\sin(\gamma_2 + \delta - \alpha - \delta_1) - \gamma_1 b\cos\delta_1}{a\cos(\alpha + \delta_1) + b\cos\delta_1} \end{bmatrix}$ |
| Type(1,2) |
| $\dot{\xi} = \begin{bmatrix} v\cos(\theta + \gamma_2 + \delta) \\ v\sin(\theta + \gamma_2 + \delta) \\ \dfrac{v\tan(\gamma_1 + \delta_1)\cos(\gamma_2 + \delta)}{a} - \dfrac{v\sin(\gamma_2 + \delta)}{a} \end{bmatrix}$ |

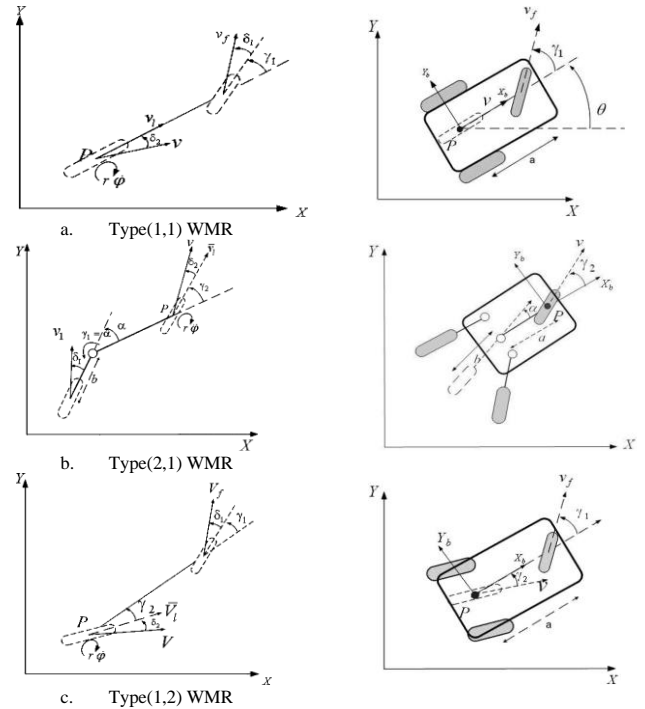

a. Type(1,1) WMR

b. Type(2,1) WMR

c. Type(1,2) WMR

Fig. 5: Different types of WMR kinematics with slippage [10]

## B. Road Tire Interaction model

Another way to come up with the slip model is to develop the road tire interaction model. As this model is developed for a wheel's motion so it can be used for a vehicle and for a WMR as well. In this way of modeling slip the effect of all the traction and resistive forces is calculated while tire's interaction with the ground during its motion. The model we are going to discuss here has been taken from [12], [13]. Fig. 6 shows resistive and tractive forces acting on the wheel during its motion. The traction force $F_t$ resulting from all the applied forces responsible for motion of the wheel is as given below.

$$F_t = F_l + F_v + F_r + F_a + F_e \quad (9)$$

Here $F_l$ is wheel inertia force and $F_v$ is vehicle inertia force, $F_r$ is rolling resistance force and $F_a$ is the cohesion force between road and tire while $F_e$ represents all the other resistive forces.

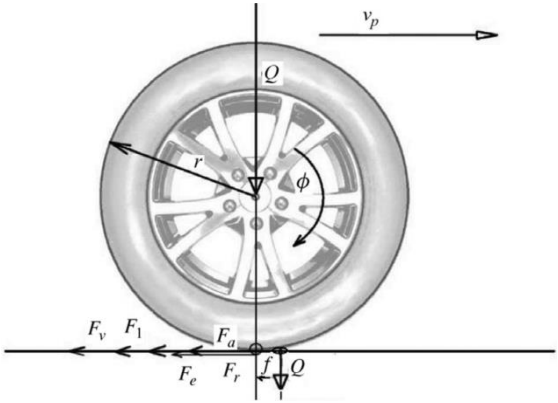

Fig. 6: Road tire interaction Model [12]

Vehicle inertia force for a robot of mass $m$ and velocity $v_p$ is given by $F_v = m\dot{v}_p$ while the rolling resistance force of wheel with radius $r$ and the force acting on the shaft is $Q$ then $F_r = f_0 Q$ where $f_0 = f/r$ is the rolling resistance coefficient with no dimensions. Cohesion force $F_a = \mu_o Q$ where $\mu_o$ is termed as wheel ground cohesion coefficient. Now substituting the values of these forces in eq. (9) we get

$$F_t = F_l + m\dot{v}_p + f_o Q + \mu_o Q + F_e \quad (10)$$

Actuation torque $M_t$ balancing the traction force on the wheel, where wheel inertial force $F_l$ will be balanced by the $j\ddot{\phi}$ while torque due to $F_v$ will be $m\dot{v}_p r$ and torque due to rolling resistance force is $f_o Qr$. Similarly $\mu_o Qp$ is the torque overcoming cohesion force where $p$ is the radius of road ground interaction point. So from eq. (10) we can deduce that

$$M_t = J\ddot{\phi} + m\dot{v}_p r + f_o Q_r + \mu_o Qp + M_e \quad (11)$$

In case of slippage actuating torque $M_t$ must be equal to $\mu_c Qr$, where $\mu_c$ is the virtual slip constant. So eq. (11) become

$$J\ddot{\phi} + m\dot{v}_p r + f_0 Qr + \mu_o Qp + M_e = \mu_c Qr \quad (12)$$

For no slip condition the robot velocity $v_p = r\dot{\phi}$, putting the value of $v_p$ in eq. (12) yields no slip condition for the wheel in motion,

$$\ddot{\phi} \leq \frac{\mu_c Qr - f_0 Qr - \mu_0 Qp - M_e}{J + mr^2} \quad (13)$$

Neglecting torque due to cohesion and other resistive forces in eq. (13) for simplicity during slip reduction, we get

$$\ddot{\phi} = \frac{\mu_c Qr}{J + mr^2} \quad (14)$$

The virtual slip coefficient $\mu_c$ must depend on slip ration $\lambda$ exponentially. So $\mu_c$ can be defined recursively as

$$\mu_c = \mu_c \exp\left(-\frac{\lambda^2}{b}\right) \quad (15)$$

Thus from eq. (14 & 15) we get

$$J\ddot{\phi} + m\dot{v}_p r = \mu_c \exp\left(-\frac{\lambda^2}{b}\right) Qr \quad (16)$$

From the slip model in eq. (3) robot velocity can be interpreted in terms of slip ratio and wheels angular velocity as

$$v_p = \dot{\phi} r(1 - \lambda(t))$$
$$\dot{v}_p = \ddot{\phi}(1 - \lambda(t)) - \dot{\phi}\dot{\lambda}(t) \quad (17)$$

Thus placing value of $\dot{v}_p$ in eq. (x), we obtain a model in terms of slip ratio $\lambda$.

$$j\ddot{\phi} + mr^2\left(\ddot{\phi}(1 - \lambda(t)) - \dot{\phi}\dot{\lambda}(t)\right) = T_r \quad (18)$$

Where $T_r = \mu_c \exp\left(-\frac{\lambda^2}{b}\right) Qr$ is the real torque induced due to the wheel slip. Thus slip in wheel can be removed by producing an equal and opposite torque to balance the torque $T_r$.

## IV. REVIEW ON SLIP AND SKID CONTROL

In last five or six years the focus of research on WMRs has tilted towards it's the violation of its nonholonomic constraints. Researchers around the world have developed control mechanism to minimize slippage considering various scenarios for diverse tasks using different control methodologies. Just recently many slip control techniques for WMRs have been devised for motion on unfavorable circumstances, path tracking and formation control. These control techniques involve kinematic and dynamics based slip control, robust control techniques like sliding mode control, model predictive control etc. In this part of the paper we are going to present a brief review of what and how of slippage control.

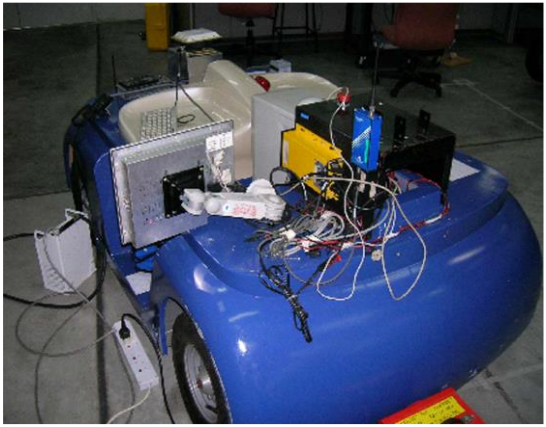

Fig. 7: RTK-GPS based kinematics model for slippage control [14]

In 2008, Low and Wang [14] controlled the motion of a WMR in the presence of slip and skid. They used Real Time Kinematics (RTK) GPS for measuring different parameters of robots motion like posture, velocities and perturbations along with some other sensors. These parameters were used by the tracking control mechanism to compensate slippage during the robot's motion. This control mechanism was implemented on a Type(1,1) WMR as shown in the Fig. 7. Experimental results validate the control mechanism.

Another control methodology for trajectory tracking of a WMR in the presence of sliding effects has been adopted in [15]. These sliding effects violate the nonholonomic constraints of the WMR's dynamic model making it highly nonlinear and time varying enforcing a need for robust control for its stabilization. Thus a second order sliding mode controller has been used for the control of the WMR. Simulation gives satisfactory tracking results.

Tarakameh *et. al.* proposed an adaptive controller in a trajectory tracking problem [16]. The proposed controller can reject the kinematic disturbances along with skid. This adaptive controller has the full parameters of the actuator in the WMR kinematics unlike other control techniques. Back stepping control methodology has be adopted has been used considering WMR as the subsystem of kinematics, dynamics and drive levels. The performance of the controller has benn demonstrated using simulation results.

Dongkyong Chwa used a robust control technique for tracking control of WMRs [17]. He proposed a new sliding mode control technique for WMRs based on kinematics in two dimensional polar co-ordinates. This technique uses different controllers for asymptotic stabilization of tracking position error and heading direction in the WMR. As per results, the controller has successfully forced the WMR to follow arbitrary paths like a circle or a straight line.

Another novel method for slip and skid control of an AGV has been proposed in [18] based on Doppler based radar sensor for measuring ground speed of the vehicle accurately. A correct estimation of friction and traction forces on tire while interacting with ground can allow a better control but this is very hard to achieve in real world. Doppler based radar along with an accelerometer has been used to measure the slip rate. Based on the measured slip rate and motor torques soil parameters are estimated and a slip control mechanism has been implemented.

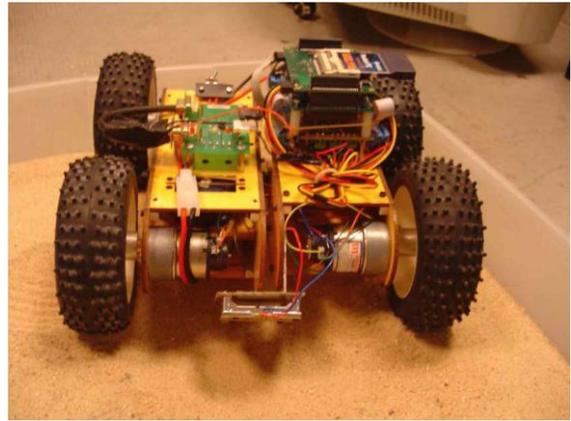

Fig. 8: Skid steering demonstration [18]

Tian and sarkar [19] studied the formation control of WMRs subject to wheel slip. However they assumed that the ideal surface can offer large amount of friction highly unlikely in practical situation. In order to investigate the effect of slip on formation control they modeled the slip-traction characteristics and integrated them into the WMR dynamics. The output linearization technique have been used to stabilize all the robots undergoing slippage.

Paulo and Urbano [20] propose a new control methodology based on kalman filter for a path following WMR with uncertainties. They presented a kalman based active observer controller to reject uncertainties and disturbances due to feedback linearization or slippage.

An adaptive skid control mechanism has been presented in [21] for stability and tracking of a vehicle in the presence of slippage. Two control schemes have been devised based on different circumstances i.e. when no road conditions are available and when some information about instant type of road surface is present. The control law manages the speed of the wheel adaptively by providing more power to drive wheel where the skid has occurred to compensate the lost driving force due to skidding. The stability analysis proved that the position and velocity errors for the vehicle are bounded.

Amodeo *et. al.* used a second order sliding mode controller for the slip and skid control of WMR during skid braking and spin acceleration[22]. The designed controller is coupled with a suitable sliding mode observer to estimate the road-tire adhesion coefficient. The antiskid braking and anti-spin acceleration has been enabled due to achieving slip's optimal value.

Just recently, Ćirović and Aleksendrić have presented an adaptive fuzzy logic based slip control for a WMR [23]. They are also involved in presenting another novel technique based on neural network for longitudinal slip control in WMRs [24].

CONCLUSION

The ever increasing applications have amplified the significance of WMRs. WMRs enjoy many inherent advantages over other type of mobile robots because of their stability and easy of use. However there are few limitations in their motion due the violation of nonholonomic constraints i.e. slip and skid. Earlier a lot of work on control of WMRs has not considered the slip and skid. That's why the performance of those control mechanisms is very poor practically. In last decade or so some work considering slip and skid in the robot dynamics has been reported for the control of WMRs. In this paper a brief overview of that work has been presented. Control of WMRs is a work in progress. The dream of watching WMRs move on slippery uneven and declining terrain is a long way away but right progress has been made in the recent past. Future of WMRs is very promising as their applications reside everywhere from home assistance to industrial application and from medicine to military.

REFERENCES

1. J. M. Yang, J. H. Kim, "Sliding mode control of trajectory tracking of nonholonomic wheeled mobile robots", IEEE Trans. Robot. Autom., 1999, 15, (3), pp. 578–587.
2. B. S. Park, S. J. Yoo, J. B. Park, H. Choiy, "Adaptive neural sliding mode control of nonholonomic wheeled mobile robots with model uncertainty", IEEE Trans. Control Syst. Technol., 2009, 17, (1), pp. 207–214.
3. T. Fukao, H. Nakagawa, N. Adachi, "Adaptive tracking control of a nonholonomic mobile robot", IEEE Trans. Robot. Autom., 2000, 16, (5), pp. 609–615.
4. W. E. Dixon, M. S. Queiroz, D. M. Dawson, T. J. Flynn, "Adaptive tracking and regulation of a wheeled mobile robot with controller/update law modularity", IEEE Trans. Control Syst. Technol., 2004, 12, (1), pp. 138–147.
5. W. Dong, K.D. Kuhnert: "Robust adaptive control of nonholonomic mobile robot with parameter and nonparameter uncertainties", IEEE Trans. Robot., 2005, 21, (2), pp. 261–266
6. R. Balakrishna and A. Ghosal, "Modeling of slip for wheeled mobile robots," IEEE Trans. on Robotics and Automation, vol. 11(1), pp. 126-132, 1995.
7. O. A. Ani, H. Xu and G. Zhao, "Analysis and modeling of slip for a five-wheeled mobile robot (WMR) in an uneven terrain," IEEE Int. Conf. on Mechatronics and Automation (ICMA), Beijing, China, 2011,pp. 154-159.
8. B. Siciliano and O. Khatib, "Wheeled Robots," in Springer handbook of robotics, Berlin, Germany, Springer, 2008, ch. 17, pp. 391-410.
9. D. Wang and C. B. Low, "Modeling skidding and slipping in wheeled mobile robots: control design perspective," IEEE/RSJ Int. Conf. Intelligent Robots and Systems (IROS), Beijing, China, 2006, pp. 1867-1872.
10. D. Wang and C. B. Low, "Modeling and analysis of skidding and slipping in wheeled mobile robots: Control design perspective," IEEE Trans. on Robotics, vol. 24(3), pp. 676-687, 2008.
11. G. Campion, G. Bastin and B. G. Andrea-Novel, "Structural properties and classification of kinematic and dynamic model of wheeled mobile robot," IEEE Trans. on Robotics and Automation, vol. 12(1), pp. 47-62, 1996.
12. T. Zielinska and A. Chmielniak, "Synthesis of control law considering wheel ground interaction and contact stability of wheeled mobile robots," Robotica Cambridge journalne, vol. 29, pp. 981-990, 2011.
13. T. Zielinska and A. Chmielniak, "Controlling the Slip in Mobile Robots," IEEE Robotics and Automation Magazine, 2009.
14. C. B. Low and D. Wang, "GPS-based path following control for a car-like wheeled mobile robot with skidding and slipping," IEEE Trans. on Control Systems Technology, vol. 16(2), pp. 340-347, 2008.
15. F. Hamerlain, K. Achour, T. Floquet and W. Perruquetti, "Higher order sliding mode control of wheeled mobile robots," IEEE Int. Conf. on Decisions and control, Seville, Spain,2005 pp. 1160-1963.
16. A. Tarakameh, "Adaptive control of nonholonomic wheeled mobile robot in the presence of lateral slip and dynamic uncertainities," IEEE Iranian conference of Electrical Engineering, Tehran, Iran, 2010.
17. D. Chwa: 'Sliding-mode control of nonholonomic wheeled mobile robots in polar coordinates', IEEE Trans. Control Syst. Technol., 2004, 12, (4), pp. 637–644
18. D. L. Desgas, C. Grand, F. B. Amer and G. C. Guiner, "Doppler based ground speed sensor fusion and slip and control for a wheeled rover," IEEE/ASME Trans. on Mechatronics, vol. 14(14), pp. 484-492, 2009.
19. Y. Tian and N. Sarkar, "Formation control of mobile robots subject to wheel slip," IEEE Int. Conf. on Robotics and Automation (ICRA), St. Paul, USA 2012pp. 4553-4558.
20. P. Coehlo, C. Nunes, "Path following control of mobile robots in presence of uncertainties," IEEE Trans. on Robotics, vol. 21(2), pp. 252-261, 2005.
21. E. F. Kececi and G. Tao, "Adaptive vehicle skid control," Elseveir J. of Mechatronics, vol. 16(5), pp. 291-301, 2006.
22. M. Amodeo, A. Ferrara, R. Terzaghi, and C. Vecchio, "Wheel slip control via second-order sliding-mode generation," IEEE Trans. on Intelligent Transportation Systems, vol. 11(1), pp. 122-131, 2010.
23. V. Ćirović, and D. Aleksendrić, "Adaptive neuro-fuzzy wheel slip control," Elsevier J. of Expert Systems with Applications, March, 2013 (In press).
24. V. Ćirović, D. Aleksendrić and D. Smiljanić, "Longitudinal wheel slip control using dynamic neural networks," Elsevier J. of Mechatronics, vol. 23(1), pp. 135-146, 2013.